\documentclass[12pt]{article}
\usepackage{axodraw,a4wide,epsfig}
\catcode`@=11
\newcount\@tempcntc
\def\@citex[#1]#2{\if@filesw\immediate\write\@auxout{\string\citation{#2}}\fi
  \@tempcnta\z@\@tempcntb\m@ne\def\@citea{}\@cite{\@for\@citeb:=#2\do
    {\@ifundefined
       {b@\@citeb}{\@citeo\@tempcntb\m@ne\@citea\def\@citea{,}{\bf ?}\@warning
       {Citation `\@citeb' on page \thepage \space undefined}}%
    {\setbox\z@\hbox{\global\@tempcntc0\csname b@\@citeb\endcsname\relax}%
     \ifnum\@tempcntc=\z@ \@citeo\@tempcntb\m@ne
       \@citea\def\@citea{,}\hbox{\csname b@\@citeb\endcsname}%
     \else
      \advance\@tempcntb\@ne
      \ifnum\@tempcntb=\@tempcntc
      \else\advance\@tempcntb\m@ne\@citeo
      \@tempcnta\@tempcntc\@tempcntb\@tempcntc\fi\fi}}\@citeo}{#1}}
\def\@citeo{\ifnum\@tempcnta>\@tempcntb\else\@citea\def\@citea{,}%
  \ifnum\@tempcnta=\@tempcntb\the\@tempcnta\else
   {\advance\@tempcnta\@ne\ifnum\@tempcnta=\@tempcntb \else \def\@citea{--}\fi
    \advance\@tempcnta\m@ne\the\@tempcnta\@citea\the\@tempcntb}\fi\fi}
\catcode`@=12

\begin{document}
\newcommand{\be}{\begin{equation}}
\newcommand{\ee}{\end{equation}}
\newcommand{\bfm}[1]{\mbox{\boldmath$#1$}}
\newcommand{\bff}[1]{\mbox{\scriptsize\boldmath${#1}$}}
\newcommand{\al}{\alpha}
\newcommand{\bt}{\beta}
\newcommand{\lm}{\lambda}
\newcommand{\bea}{\begin{eqnarray}}
\newcommand{\eea}{\end{eqnarray}}
\newcommand{\gm}{\gamma}
\newcommand{\Gm}{\Gamma}
\newcommand{\dl}{\delta}
\newcommand{\Dl}{\Delta}
\newcommand{\ep}{\epsilon}
\newcommand{\vep}{\varepsilon}
\newcommand{\kp}{\kappa}
\newcommand{\Lm}{\Lambda}
\newcommand{\om}{\omega}
\newcommand{\pa}{\partial}
\newcommand{\nn}{\nonumber}
\newcommand{\dd}{\mbox{d}}
\newcommand{\grtsim}{\mbox{\raisebox{-3pt}{$\stackrel{>}{\sim}$}}}
\newcommand{\lessim}{\mbox{\raisebox{-3pt}{$\stackrel{<}{\sim}$}}}
\newcommand{\uk}{\underline{k}}
\newcommand{\gsim}{\;\rlap{\lower 3.5 pt \hbox{$\mathchar \sim$}} \raise 1pt \hbox {$>$}\;}
\newcommand{\lsim}{\;\rlap{\lower 3.5 pt \hbox{$\mathchar \sim$}} \raise 1pt \hbox {$<$}\;}
\newcommand{\Li}{\mbox{Li}}
\newcommand{\bc}{\begin{center}}
\newcommand{\ec}{\end{center}}

\begin{titlepage}
\nopagebreak
{\flushright{
        \begin{minipage}{5cm}
         ALBERTA-THY-18-07 \\
         IFIC/07-66 \\
         ZU-TH 26/07 \\
        {\tt hep-ph/yymmnnn}
        \end{minipage}        }

}
\renewcommand{\thefootnote}{\fnsymbol{footnote}}
\vskip 3cm
\begin{center}
\boldmath
{\Large\bf Two-Loop Heavy-Flavor Contribution  \\[7pt]
to Bhabha Scattering}\unboldmath
\vskip 1.0cm
{\large  R.~Bonciani$\rm \, ^{a, \,}$\footnote{Email: {\tt
Roberto.Bonciani@ific.uv.es}}},
{\large A.~Ferroglia$\rm \, ^{b, \,}$\footnote{Email:
{\tt Andrea.Ferroglia@physik.unizh.ch}}},
and  {\large A.A.~Penin$\rm \, ^{c, \, d, \,}$\footnote{Email: {\tt apenin@phys.ualberta.ca}}}
\vskip .7cm
{\it $\rm ^a$ Departamento de F\'{\i}sica Te\`orica,
IFIC, CSIC -- Universidad de
Valencia, \\
E-46071 Valencia, Spain}
\vskip .3cm
{\it $\rm ^b$ Institut f{\"u}r Theoretische Physik,
Universit{\"a}t Z\"urich, \\
CH-8057 Zurich, Switzerland}
\vskip .3cm
{\it $\rm ^c$ Department of Physics, University Of Alberta, \\
Edmonton, AB T6G 2J1, Canada}
\vskip .3cm
{\it $\rm ^d$ Institute for Nuclear Research of Russian Academy of Sciences, \\
117312 Moscow, Russia}
\end{center}
\vskip .5cm

\begin{abstract}
  We evaluate the two-loop QED corrections to the
  Bhabha scattering cross section which involve
  the vacuum polarization by heavy
  fermions of arbitrary mass $m_f\gg m_e$. The  results are valid
  for generic values of the Mandelstam invariants $s,t,u \gg m_e^2$.
\\[2mm]
\end{abstract}
\vfill
\end{titlepage}


\section{Introduction}

\label{int} Electron-positron {\it Bhabha} scattering \cite{Bha}
plays a special role in particle phenomenology. It provides a very
efficient tool for luminosity determination at electron-positron
colliders and thus it is crucial for extracting physics from the
experimental data.  Small angle Bhabha scattering is particularly
effective as a luminosity monitor at the high energy colliders such
as LEP and the future International Linear Collider (ILC)
\cite{Jad,MNP}. The large angle Bhabha scattering is used to measure
the luminosity at colliders operating at a center of mass energy
$\sqrt{s}$ of a few GeV, such as BABAR/PEP-II, BELLE/KEKB, BES/BEPC,
KLOE/DA$\Phi$NE, and VEPP-2M \cite{Car}. The large angle Bhabha
scattering will also be employed in order to disentangle the
luminosity spectrum at ILC \cite{Too,Heu}. Since the accuracy of the
theoretical evaluation of the Bhabha cross section directly affects
the luminosity determination, remarkable efforts have been devoted
to the study of the radiative corrections to this process (see
\cite{Jad} for an extensive list of references).  QED contributions
dominate the radiative corrections to the large angle scattering at
intermediate energies 1-10~GeV and to the small angle scattering
also at higher energies.  The calculation of the QED radiative
corrections to the Bhabha cross section is among the classical
problems of perturbative quantum field theory and has a long
history.  The first-order corrections are well known (see \cite{Ber}
and references therein).  However, to match the impressive
experimental accuracy reached at colliders, the complete
second-order QED effects have to be taken into account. The
evaluation of the two-loop virtual corrections is the main problem
of the second-order analysis. The two-loop QED corrections can be
divided into three main categories: (i) the corrections involving
the electron vacuum polarization {\it i.e.} with at least one closed
electron loop, (ii) the pure photonic corrections, and (iii) the
corrections involving the vacuum polarization by heavy flavors
(leptons or quarks) of  mass $m_f\gg m_e$. Type (i) corrections have
been evaluated in \cite{BMR-V2} including the full dependence on the
electron mass $m_e$.  For the virtual corrections of type (ii) the
full dependence on the electron mass is known with the exception of
the two-loop box diagrams \cite{BF}.\footnote{Partial results for
the massive Bhabha scattering two-loop box
 diagrams are discussed in \cite{Heinrich:2004iq,Smirnov:2001cm}.}
At the same time, the complete result for  the pure photonic
corrections has been obtained in \cite{Pen} in the leading order of
the small electron mass expansion through the {\it infrared
matching} to the massless approximation \cite{BDG}.\footnote{The
terms enhanced by logarithms of the electron mass have been derived
in this approximation in \cite{ArbPap,Bas}.} It is sufficient for
all the phenomenological applications and has been recently
confirmed within a slightly different framework \cite{BecMel}. For
the heavy-flavor contribution the result is available only in the
limit $m_f^2 \ll s,t,u $ \cite{BecMel,Act}. This condition, however,
does not hold for $\tau$-lepton, $c$- and $b$-quark in the
practically interesting energy range of about a few GeV as well as
for the top quark at the typical ILC energies $500~{\rm
GeV}\lsim\sqrt{s}\lsim 1000~{\rm GeV}$.

In this letter we consider the two-loop heavy-flavor contribution
retaining the full dependence on $m_f^2/s$. In the next section we
outline the structure of the corrections   and the calculation method.
In Sect.~\ref{sec3} we present the complete numerical result for the
correction in two phenomenologically interesting cases.  Sect.~\ref{sum}
contains our conclusions.  The expansion of the result in the limits
$m_f^2 \ll s$ and $m_f^2 \gg s$ is given in the Appendix in analytic form.

\section{Structure of the Radiative Corrections and Calculation Method
\label{sec2}}

We consider the phenomenologically interesting kinematical region
$s,~t,~u\sim m_f^2\gg m_e^2$, where all the terms suppressed by the
electron mass can be neglected. The perturbative expansion for the
Bhabha cross section in the fine structure constant $\al$ is defined as
follows
\begin{equation}
{{\rm d}\sigma\over {\rm d}\Omega}=\sum_{n=0}^\infty
\left({\al\over\pi}\right)^n
\delta^{(n)}{{\rm d}\sigma^{(0)}\over {\rm d}\Omega}\,,
\label{sigser}
\end{equation}
where $\delta^{(n)}$ is the correction factor, with $\delta^{(0)}=1$. The
leading order differential cross section  in the small electron mass
approximation takes the form
\begin{equation}
{{\rm d}\sigma^{(0)}\over{\rm d}\Omega} ={\al^2\over
s}\left({1-x+x^2\over x}\right)^2+{\mathcal O}(m_e^2/s)\,,
\label{losig}
\end{equation}
where $x=(1-\cos\theta)/2$ and $\theta$ is the scattering angle. The
virtual corrections are infrared divergent. These {\it soft}
divergencies are canceled in the inclusive cross section when one
adds the photonic bremsstrahlung \cite{Kin}.  The standard approach
to deal with the bremsstrahlung is to split it into a soft part,
which accounts for the emission of the photons with the energy below
a given cutoff $\vep_{cut}\ll m_e$, and a hard part corresponding to
the emission of the photons with the energy larger than
$\vep_{cut}$.  The infrared finite hard part is then computed
numerically using Monte-Carlo methods with physical cuts dictated by
the experimental setup.  At the same time, the soft part is computed
analytically and combined with the virtual corrections ensuring the
cancellation of the soft divergencies in Eq.~(\ref{sigser}). We
regulate all the soft divergencies by dimensional regularization in
$d=4-2\vep$ space-time dimensions.

The first-order heavy-flavor contribution  to the cross section reads
\be
\delta^{(1)} = \frac{1}{1-x+x^2}\biggl\{ x (2 x -1) {\rm
    Re}\!\left[\Pi_f^{(1)}(\rho)\right] + (2-x)
    {\rm Re}\!\left[\Pi_f^{(1)}(\rho/x)\right] \biggr\} \, ,
\label{1ldec}
\ee
where
\be
\Pi_f^{(1)}(\rho) = {Q_f^2 N_c\over 3} \left[-{5\over 3} + 4\rho
- \left( 1+2 \rho \right) \sqrt{1-4\rho}
\left( 2 \, {\rm  arctanh} \! \sqrt{\frac{1}{1\!-\!4\rho}}
+i\pi\,\theta \left( \frac{1}{\rho}-4 \right) \right) \right] \, ,
\label{1lpi}
\ee
is the well known one-loop vacuum polarization function,
$\rho=m_f^2/s$, $Q_f$ is the heavy-flavor electric charge, and
the number of colors $N_c$ is 1 for leptons and
3 for quarks. We
work within the standard on-shell QED renormalization scheme.  The
second-order contribution can be split in the sum of two terms
\be
\delta^{(2)} = \delta^{(2)}_{vv}+\delta^{(1)}_{s}\delta^{(1)} \, ,
\label{2ldec}
\ee
which correspond to the two-loop virtual correction\footnote{We do not
  consider the trivial correction given by two heavy-fermion loop
  insertions.} and the one-loop correction to the single soft photon
emission which factorizes into the product of the first-order
contributions \cite{YFS}.

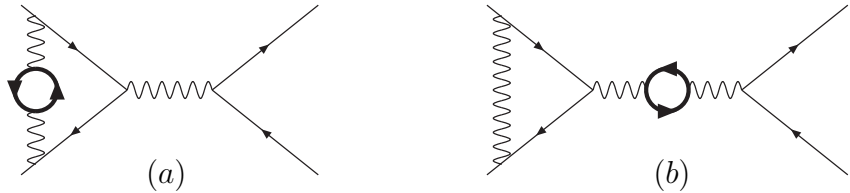
\begin{figure}[t]
\begin{center}
\hspace*{10mm}
\begin{picture}(140,80)(0,0)
\SetScale{0.8}
\SetWidth{0.6}
\ArrowLine(90,40)(140,80)
\ArrowLine(140,0)(90,40)
\ArrowLine(50,40)(0,0)
\ArrowLine(0,80)(50,40)
\Photon(7,5)(7,30){4}{3.5}
\Photon(7,50)(7,75){4}{3.5}
\Photon(50,40)(90,40){4}{5.5}
\SetWidth{2}
\ArrowArc(7,40)(10,90,270)
\ArrowArc(7,40)(10,270,90)
\Text(55,0)[cc]{$(a)$}
\end{picture}
\hspace*{10mm}
\begin{picture}(170,80)(0,0)
\SetScale{0.8}
\SetWidth{0.6}
\ArrowLine(120,40)(170,80)
\ArrowLine(170,0)(120,40)
\Photon(95,40)(120,40){4}{3.5}
\ArrowLine(50,40)(0,0)
\ArrowLine(0,80)(50,40)
\Photon(7,5)(7,75){4}{10.5}
\Photon(50,40)(75,40){4}{3.5}
\SetWidth{2}
\ArrowArc(85,40)(10,0,180)
\ArrowArc(85,40)(10,180,0)
\Text(70,0)[cc]{$(b)$}
\end{picture}
\end{center}
\caption{\label{fig1} \small The
  two-loop diagrams associated with the logarithmic dependence of the
  corrections to the Bhabha scattering amplitude on $m_e$. Actually the
  diagram $(a)$ is free of electron mass logarithms. The bold arrow
  circle corresponds to the heavy-flavor vacuum polarization. }
\end{figure}

The calculation of the virtual corrections is a highly nontrivial
problem since it involves the two-loop four-point Feynman integrals
that depend on four independent mass parameters. However, the calculation  is
significantly simplified in the small electron mass limit.
Eq.~(\ref{2ldec}) can be rewritten according to the asymptotic
dependence on $m_e$
\begin{equation}
\delta^{(2)} = \delta^{(2)}_{1}\ln\left({s\over m_e^2} \right) +
\delta^{(2)}_{0} + {\mathcal O}(m_e^2/s)\, . \label{2lexp}
\end{equation}
The logarithmic term in Eq.~(\ref{2lexp}) is a remnant of the {\it
  collinear} divergence which is regulated by the electron mass.
 The quantities $ \delta^{(2)}_{1}$ and $\delta^{(2)}_{0}$ in
Eq.~(\ref{2lexp}) depend on $s$, $t$, and $m_f$ only. The
collinear divergencies and hence the singular dependence of the
corrections on $m_e$ have a peculiar structure which was
extensively studied in the context of QCD. In particular, the
collinear divergencies factorize and can be absorbed in the
external field renormalization \cite{FreTay}. This property is
crucial for our analysis because it allows us to perform the most
difficult part of the calculation with a strictly massless
electron.  Indeed, due to the factorization, the singular
dependence on $m_e$ is the same for the Bhabha amplitude and (the
square of) the vector form factor \cite{Pen}. This attributes the
total logarithmic corrections to the two-loop Bhabha scattering amplitude
to  the diagrams shown in Fig.~\ref{fig1}. However, due to the
renormalization condition the vacuum polarization does not change
the photon propagator near the mass shell where the collinear
divergencies are located. As a result, the diagram $(a)$ is
infrared finite even for $m_e=0$ and the singular terms are
entirely contained in the one-particle reducible diagram $(b)$.  In calculating
 the cross section one must take into account also the contributions coming
from the interference of the one-loop corrections to the amplitude and
the corresponding soft emission; both contributions have a factorized form.
Thus it is straightforward to obtain the coefficient of the logarithmic term in
Eq.~(\ref{2lexp}) which reads
\begin{equation}
\delta^{(2)}_{1} = \left[ 4\ln\left({\vep_{cut}\over\vep}\right) + 3
\right] \delta^{(1)} \, , \label{del21}
\end{equation}
where $\vep=\sqrt{s}/2$.  At the same time the sum of the
one-particle irreducible diagrams has a regular behavior in the
small electron mass limit and can be computed with $m_e=0$.  Let us
emphasize that this property in general holds only for the sum of
the diagrams.  The individual diagrams computed in a covariant gauge
do exhibit the collinear divergencies for $m_e=0$.  This however
does not pose any additional problem since we work in dimensional
regularization. In this case the collinear divergencies show up as
the extra poles in $\vep$ which are not related to the soft emission
and disappear in the sum of the one-particle irreducible diagrams.
Thus all the ``true'' two-loop diagrams contribute only to the non
logarithmic term in Eq.~(\ref{2lexp}) which can be written as
follows
\begin{equation}
\delta^{(2)}_{0} = -4\left[1+\ln\left({1-x\over x}\right)
\right]\ln\left({\vep_{cut}\over\vep}\right)\delta^{(1)} + Q_f^2 N_c
\left({x\over 1-x+x^2}\right)^2f(\rho,x) \, , \label{del20}
\end{equation}
where the first term is determined by the soft emission and
$f(\rho,x)$ is a function of two dimensionless variables:
$\rho=m_f^2/s$ and $x=-t/s$.  The two-loop problem with massless
electron falls it the same complexity class as the one considered in
Ref.~\cite{BMR-V}. The most difficult part of the calculation is
represented, as expected, by the evaluation of the two-loop box
graphs. By employing the Laporta Algorithm \cite{Laporta} to
identify the set of master integrals (MI), it is possible to
conclude that just two of the MI necessary for this calculation
correspond to four point functions. These two are the only unknown
ones among the MIs appearing in the calculation. They were evaluated
by means of the differential equation method \cite{diffeq} and
expressed in terms of Harmonic Polylogarithms \cite{HPLs}, and
suitable generalizations of the latter \cite{AB2}. The technical
aspects of the calculation are going to be discussed in detail in
\cite{BFP2}.
Finally, the result for the function $f(\rho,x)$ can be
expressed in closed analytic form in terms of one- and two-dimensional
generalized harmonic polylogarithms of maximum weight three.

\section{Results and Numerical Estimates \label{sec3}}

The complete analytical result is rather lengthy and will be published
elsewhere. In this letter we focus on the numerical impact of the
correction in the phenomenologically important cases and we present the
approximate analytical expression in different limits.

\begin{figure}
\bc
\begin{picture}(0,0)%
\includegraphics{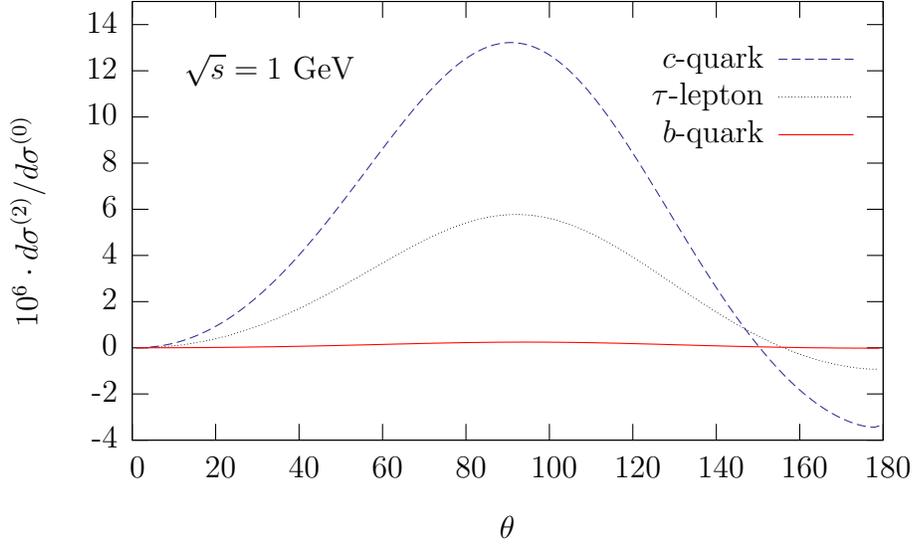}%
\end{picture}%
\begingroup
\setlength{\unitlength}{0.0200bp}%
\begin{picture}(18000,10800)(0,0)%
\put(2750,1979){\makebox(0,0)[r]{\strut{}-4}}%
\put(2750,2850){\makebox(0,0)[r]{\strut{}-2}}%
\put(2750,3720){\makebox(0,0)[r]{\strut{} 0}}%
\put(2750,4591){\makebox(0,0)[r]{\strut{} 2}}%
\put(2750,5462){\makebox(0,0)[r]{\strut{} 4}}%
\put(2750,6332){\makebox(0,0)[r]{\strut{} 6}}%
\put(2750,7203){\makebox(0,0)[r]{\strut{} 8}}%
\put(2750,8073){\makebox(0,0)[r]{\strut{} 10}}%
\put(2750,8944){\makebox(0,0)[r]{\strut{} 12}}%
\put(2750,9815){\makebox(0,0)[r]{\strut{} 14}}%
\put(3025,1429){\makebox(0,0){\strut{} 0}}%
\put(4597,1429){\makebox(0,0){\strut{} 20}}%
\put(6169,1429){\makebox(0,0){\strut{} 40}}%
\put(7742,1429){\makebox(0,0){\strut{} 60}}%
\put(9314,1429){\makebox(0,0){\strut{} 80}}%
\put(10886,1429){\makebox(0,0){\strut{} 100}}%
\put(12458,1429){\makebox(0,0){\strut{} 120}}%
\put(14031,1429){\makebox(0,0){\strut{} 140}}%
\put(15603,1429){\makebox(0,0){\strut{} 160}}%
\put(17175,1429){\makebox(0,0){\strut{} 180}}%
\put(10100,275){\makebox(0,0){\strut{} $\theta$ }}%
\put(3811,8944){\makebox(0,0)[l]{\strut{} $\sqrt{s}=1$~GeV }}%
\put(14935,9162){\makebox(0,0)[r]{\strut{}$c$-quark}}%
\put(14935,8450){\makebox(0,0)[r]{\strut{}$\tau$-lepton}}%
\put(14935,7738){\makebox(0,0)[r]{\strut{}$b$-quark}}%
\put(1100,6114){\rotatebox{90}{\makebox(0,0){\strut{} $10^6 \cdot d\sigma^{(2)}/d\sigma^{(0)}$ }}}%
\end{picture}%
\endgroup
\caption{\small{Two-loop corrections to the Bhabha scattering
differential cross section at $\sqrt{s}=1$~GeV due to a closed loop
of $\tau$-lepton (dotted line), $c$-quark (dashed line) and
$b$-quark (solid line) for $m_c=1.25$~GeV and $m_b=4.7$~GeV.}
\label{fig2}}
\ec
\end{figure}
\begin{figure}
\bc
\begin{picture}(0,0)%
\includegraphics{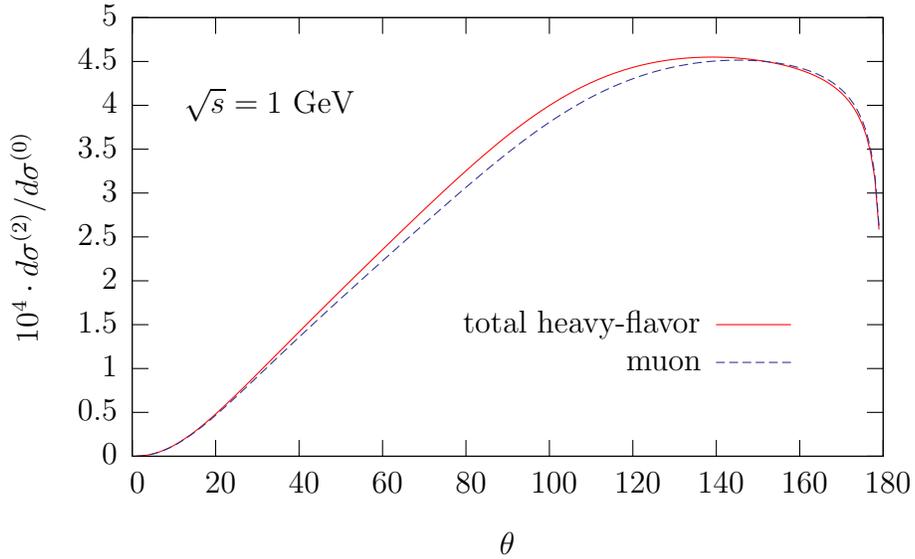}%
\end{picture}%
\begingroup
\setlength{\unitlength}{0.0200bp}%
\begin{picture}(18000,10800)(0,0)%
\put(2750,1979){\makebox(0,0)[r]{\strut{} 0}}%
\put(2750,2806){\makebox(0,0)[r]{\strut{} 0.5}}%
\put(2750,3633){\makebox(0,0)[r]{\strut{} 1}}%
\put(2750,4460){\makebox(0,0)[r]{\strut{} 1.5}}%
\put(2750,5287){\makebox(0,0)[r]{\strut{} 2}}%
\put(2750,6115){\makebox(0,0)[r]{\strut{} 2.5}}%
\put(2750,6942){\makebox(0,0)[r]{\strut{} 3}}%
\put(2750,7769){\makebox(0,0)[r]{\strut{} 3.5}}%
\put(2750,8596){\makebox(0,0)[r]{\strut{} 4}}%
\put(2750,9423){\makebox(0,0)[r]{\strut{} 4.5}}%
\put(2750,10250){\makebox(0,0)[r]{\strut{} 5}}%
\put(3025,1429){\makebox(0,0){\strut{} 0}}%
\put(4597,1429){\makebox(0,0){\strut{} 20}}%
\put(6169,1429){\makebox(0,0){\strut{} 40}}%
\put(7742,1429){\makebox(0,0){\strut{} 60}}%
\put(9314,1429){\makebox(0,0){\strut{} 80}}%
\put(10886,1429){\makebox(0,0){\strut{} 100}}%
\put(12458,1429){\makebox(0,0){\strut{} 120}}%
\put(14031,1429){\makebox(0,0){\strut{} 140}}%
\put(15603,1429){\makebox(0,0){\strut{} 160}}%
\put(17175,1429){\makebox(0,0){\strut{} 180}}%
\put(10100,275){\makebox(0,0){\strut{} $\theta$ }}%
\put(3811,8596){\makebox(0,0)[l]{\strut{} $\sqrt{s}=1$~GeV }}%
\put(13756,4460){\makebox(0,0)[r]{\strut{}total heavy-flavor}}%
\put(13756,3748){\makebox(0,0)[r]{\strut{}muon}}%
\put(1100,6114){\rotatebox{90}{\makebox(0,0){\strut{} $10^4 \cdot d\sigma^{(2)}/d\sigma^{(0)}$ }}}%
\end{picture}%
\endgroup
\caption{\small{Two-loop corrections to the Bhabha scattering
differential cross section at $\sqrt{s}=1$~GeV due to a closed loop of
muon (dashed line). The solid line represents the sum of the
contributions of the muon, $\tau$-lepton, $c$-quark and $b$-quark.}
\label{fig3}}
\ec
\end{figure}
\begin{figure}
\bc
\begin{picture}(0,0)%
\includegraphics{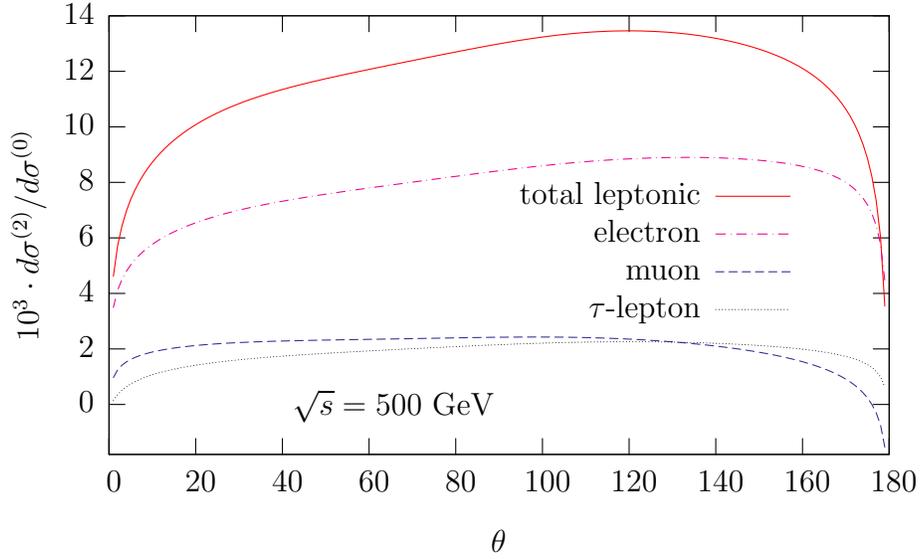}%
\end{picture}%
\begingroup
\setlength{\unitlength}{0.0200bp}%
\begin{picture}(18000,10800)(0,0)%
\put(2200,2921){\makebox(0,0)[r]{\strut{} 0}}%
\put(2200,3968){\makebox(0,0)[r]{\strut{} 2}}%
\put(2200,5015){\makebox(0,0)[r]{\strut{} 4}}%
\put(2200,6062){\makebox(0,0)[r]{\strut{} 6}}%
\put(2200,7109){\makebox(0,0)[r]{\strut{} 8}}%
\put(2200,8156){\makebox(0,0)[r]{\strut{} 10}}%
\put(2200,9203){\makebox(0,0)[r]{\strut{} 12}}%
\put(2200,10250){\makebox(0,0)[r]{\strut{} 14}}%
\put(2475,1429){\makebox(0,0){\strut{} 0}}%
\put(4108,1429){\makebox(0,0){\strut{} 20}}%
\put(5742,1429){\makebox(0,0){\strut{} 40}}%
\put(7375,1429){\makebox(0,0){\strut{} 60}}%
\put(9008,1429){\makebox(0,0){\strut{} 80}}%
\put(10642,1429){\makebox(0,0){\strut{} 100}}%
\put(12275,1429){\makebox(0,0){\strut{} 120}}%
\put(13908,1429){\makebox(0,0){\strut{} 140}}%
\put(15542,1429){\makebox(0,0){\strut{} 160}}%
\put(17175,1429){\makebox(0,0){\strut{} 180}}%
\put(9825,275){\makebox(0,0){\strut{} $\theta$ }}%
\put(5742,2869){\makebox(0,0)[l]{\strut{} $\sqrt{s}=500$~GeV }}%
\put(13633,6847){\makebox(0,0)[r]{\strut{}total leptonic}}%
\put(13633,6135){\makebox(0,0)[r]{\strut{}electron}}%
\put(13633,5423){\makebox(0,0)[r]{\strut{}muon}}%
\put(13633,4711){\makebox(0,0)[r]{\strut{}$\tau$-lepton}}%
\put(1000,6114){\rotatebox{90}{\makebox(0,0){\strut{} $10^3 \cdot d\sigma^{(2)}/d\sigma^{(0)}$ }}}%
\end{picture}%
\endgroup
\caption{\small{Two-loop leptonic corrections to the Bhabha scattering
differential cross section at $\sqrt{s}=500$~GeV. The dash-dotted
line represents the electron contribution including the soft-pair
radiation. The dashed and dotted lines represent the contributions
of muon and $\tau$-lepton, respectively. The solid line  is the sum
of the three contributions.} \label{fig4}}
\ec
\end{figure}
\begin{figure}
\bc
\begin{picture}(0,0)%
\includegraphics{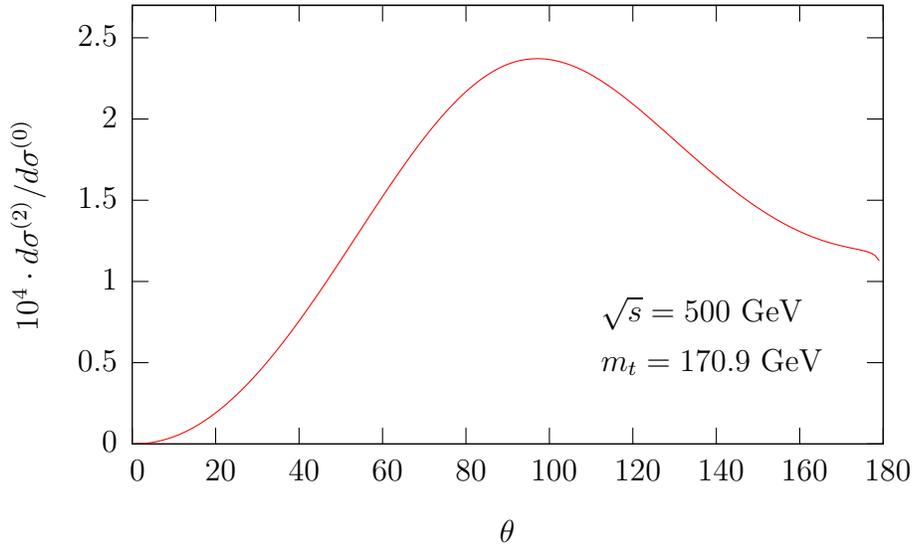}%
\end{picture}%
\begingroup
\setlength{\unitlength}{0.0200bp}%
\begin{picture}(18000,10800)(0,0)%
\put(2750,1979){\makebox(0,0)[r]{\strut{} 0}}%
\put(2750,3511){\makebox(0,0)[r]{\strut{} 0.5}}%
\put(2750,5042){\makebox(0,0)[r]{\strut{} 1}}%
\put(2750,6574){\makebox(0,0)[r]{\strut{} 1.5}}%
\put(2750,8106){\makebox(0,0)[r]{\strut{} 2}}%
\put(2750,9637){\makebox(0,0)[r]{\strut{} 2.5}}%
\put(3025,1429){\makebox(0,0){\strut{} 0}}%
\put(4597,1429){\makebox(0,0){\strut{} 20}}%
\put(6169,1429){\makebox(0,0){\strut{} 40}}%
\put(7742,1429){\makebox(0,0){\strut{} 60}}%
\put(9314,1429){\makebox(0,0){\strut{} 80}}%
\put(10886,1429){\makebox(0,0){\strut{} 100}}%
\put(12458,1429){\makebox(0,0){\strut{} 120}}%
\put(14031,1429){\makebox(0,0){\strut{} 140}}%
\put(15603,1429){\makebox(0,0){\strut{} 160}}%
\put(17175,1429){\makebox(0,0){\strut{} 180}}%
\put(10100,275){\makebox(0,0){\strut{} $\theta$ }}%
\put(11672,4430){\makebox(0,0)[l]{\strut{} $\sqrt{s}=500$~GeV }}%
\put(11672,3511){\makebox(0,0)[l]{\strut{} $m_t=170.9$~GeV }}%
\put(1100,6114){\rotatebox{90}{\makebox(0,0){\strut{} $10^4 \cdot d\sigma^{(2)}/d\sigma^{(0)}$ }}}%
\end{picture}%
\endgroup
\caption{\small{Two-loop corrections to the Bhabha scattering differential
cross section at $\sqrt{s}=500$~GeV due to a closed loop of top
quark for $m_t=170.9$~GeV. } \label{fig5}}
\ec
\end{figure}
\begin{figure}
\bc
\begin{picture}(0,0)%
\includegraphics{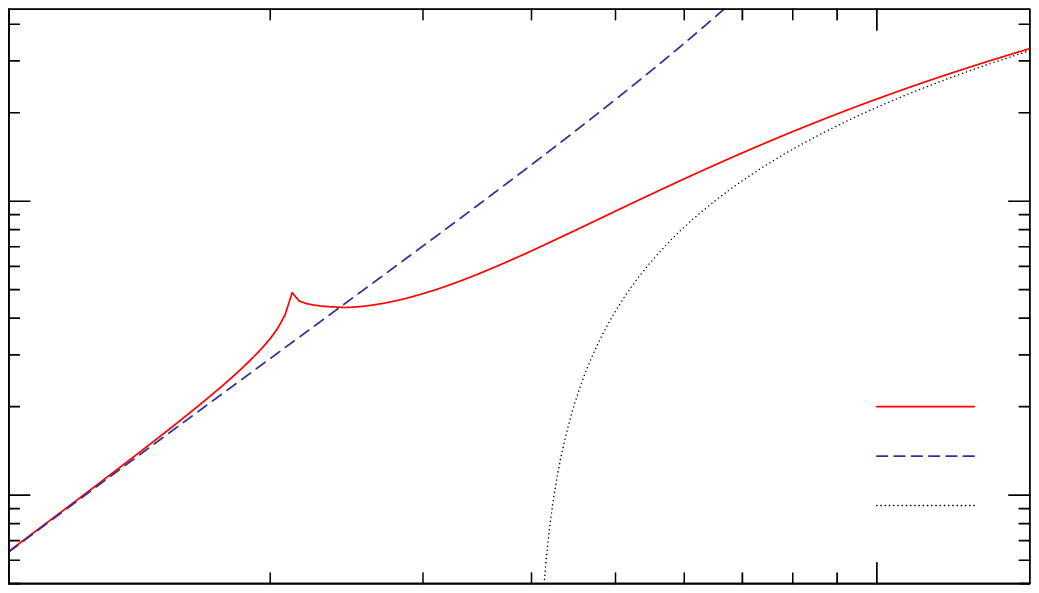}%
\end{picture}%
\begingroup
\setlength{\unitlength}{0.0200bp}%
\begin{picture}(18000,10800)(0,0)%
\put(2200,3253){\makebox(0,0)[r]{\strut{} 0.1}}%
\put(2200,7485){\makebox(0,0)[r]{\strut{} 1}}%
\put(2475,1429){\makebox(0,0){\strut{} 0.1}}%
\put(14974,1429){\makebox(0,0){\strut{} 1}}%
\put(9825,275){\makebox(0,0){\strut{} $\sqrt{s}$ (GeV)}}%
\put(3465,8759){\makebox(0,0)[l]{\strut{} $\theta=60^{\circ}$ }}%
\put(14699,4527){\makebox(0,0)[r]{\strut{}exact result}}%
\put(14699,3815){\makebox(0,0)[r]{\strut{}large-mass expansion}}%
\put(14699,3103){\makebox(0,0)[r]{\strut{}small-mass expansion}}%
\put(800,6114){\rotatebox{90}{\makebox(0,0){\strut{} $10^4 \cdot d\sigma^{(2)}/d\sigma^{(0)}$ }}}%
\end{picture}%
\endgroup
\caption{\small{ Two-loop corrections to the Bhabha scattering
differential cross section at $\theta=60^{\circ}$ due to a closed
loop of muon. The solid line represents the exact result. The dashed
and dotted lines represent  the results of the large-mass
expansion (Eq.~(\ref{f1bar}))  and  small-mass expansion
(Eq.~(\ref{f0})), respectively.} \label{fig6}}
\ec
\end{figure}

As the first application, we consider the Bhabha scattering at the
energy of 1~GeV which plays a crucial role in the determination of
the hadronic vacuum polarization contribution to the muon anomalous
magnetic moment \cite{Alo,Eid}.  The second-order contributions
$(\al/\pi)^2\delta^{(2)}$ of heavy leptons and quarks to the
differential cross section are plotted as function of the scattering
angle in Figs.~\ref{fig2} and \ref{fig3} for $\sqrt{s}=1$~GeV, which
is the center-of-mass energy of the KLOE experiment at DA$\Phi$NE.
All the terms involving a power of the logarithm
$\ln(\vep_{cut}/\vep)$ are excluded from the numerical estimates
because the corresponding contribution critically depends on the
event selection algorithm and cannot be unambiguously estimated
without imposing specific cuts on the photon bremsstrahlung.  Note
that the energy under consideration is sufficiently below the
quarkonium threshold and the heavy-quarks can be treated
perturbatively.  The contributions of the $\tau$-lepton, $c$ and $b$
quarks are suppressed with respect to the muon at least by one order
of magnitude and they are given separately in Fig.~\ref{fig2}. Thus
the total heavy-flavor contribution is dominated by the muon loop
and it reaches 0.45 permille in magnitude at $\theta\sim 140^\circ$,
see Fig.~\ref{fig3}.

Let us now discuss  Bhabha scattering at high energies
characteristic to the ILC.  We consider only the contributions of
the leptons and the top quark because  the lighter quarks need a
special treatment due to hadronization effects \cite{KKKS}. In
Fig.~\ref{fig4} we present  the plots for the contributions of muon,
$\tau$-lepton, and electron (including the soft electron-positron
pair emission) at $\sqrt{s}=500$~GeV. The total leptonic correction
reaches $1.3$\% at $\theta\sim 120^{\circ}$ and it is comparable
with the pure photonic term \cite{Pen}. The contribution of top
quark is plotted in Fig.~\ref{fig5}. It is significantly smaller
than the leptonic one.

The analytical structure of the result becomes much simpler in the
small-mass limit $m_f^2\ll s$ and the large-mass limit $m_f^2\gg s$.
The corresponding expansions of the function $f(\rho,x)$ at $\rho\to
0$ and $\rho\to\infty$ are given in the Appendix.
Both expansions break down near the threshold
$\sqrt{s}\sim 2m_f$. In Fig.~\ref{fig6}  the muon contribution at
the fixed scattering angle $\theta=60^{\circ}$ is plotted as
function of $\sqrt{s}$ in the threshold region around $2m_\mu\approx
216$~MeV. In this plot we present the exact result along with the
first nonvanishing  terms of the small and large mass expansions,
Eqs.~(\ref{f0},~\ref{f1bar}).  We observe that the asymptotic
results fail to approximate the exact one with 10\% accuracy in the
interval $1.7\lsim \sqrt{s}/m_f\lsim 10$. In particular they are
completely useless for the description of the top quark contribution
at the typical energies of the ILC.

\section{Summary \label{sum}}

To conclude, we have derived the two-loop radiative corrections to
Bhabha scattering due to the heavy-flavor vacuum polarization in the
leading order of the small electron mass expansion and for any ratio
of the heavy-fermion mass to the  energy of the process. This
completes the calculation of the QED part of the two-loop
corrections. We have analyzed the phenomenological impact of our
result in two different experimental setups. For  $e^{+}e^{-}$
colliders operating at the  energies of a few GeV, the correction is
dominated by  the virtual  muon loop  and reaches 0.45 permille at
large scattering angles. At the typical ILC energies the
$\tau$-lepton and muon loops result in the comparable  contributions
of about two permille  so that  the total leptonic correction amounts
up to 1.3\%. The contribution of the top quark is smaller and at
$\sqrt{s}=500$~GeV reaches approximately  0.25 permille in magnitude.

\subsection*{Acknowledgements}

We are grateful to J.~Vermaseren and D. Maitre for their kind assistance in the use of
{\tt FORM} \cite{FORM}, and of the {\tt Mathematica} packages {\tt HPL} and
{HypExp} \cite{Daniel1}. A.~F. and R.~B. would like to thank S.~Actis
for useful discussions. R.~B. would like to thank D. Greynat for useful discussions
about Mellin-Barnes \cite{david}, the Galileo Galilei Institute for Theoretical
Physics for the hospitality and the INFN for partial support during the completion
of this work. The work of R.~B. was partially supported by Ministerio de Educaci\'on
y Ciencia (MEC) under grant FPA2004-00996, Generalitat Valenciana under grant
GVACOMP2007-156, European Commission under the grant MRTN-CT-2006-035482 (FLAVIAnet),
and MEC-INFN agreement.
The work of A.~F. was supported
in part by the Swiss National Science
Foundation (SNF) under contract 200020-117602.
The work of A.~P. was supported in part by BMBF Grant No.\ 05HT4VKA/3 and
Sonderforschungsbereich Transregio 9.

\section*{Appendix}
The small-mass expansion of the function $f(\rho,x)$ is of the following
form
\begin{equation}
f(\rho,x)=\sum_{n=0}^\infty\rho^nf_n(\rho,x)\, ,
\label{smexp}
\end{equation}
where $f_n(\rho,x)$ depend on $\rho$ only logarithmically.
For the  leading term we obtain
\begin{eqnarray}
& & \hspace{-5mm} f_0(\rho,x) = \frac{\left(x^2-x+1\right)^2}{x^2}\Biggl\{
\frac{1}{9} \ln^3\left(\rho\right)
+ \ln^2\left(\rho\right)\Biggl[\frac{1}{3} \ln(1 - x)
+ \frac{19 }{18}
-\frac{1}{3} \ln(x) \Biggr]\nn\\
& & \hspace{10mm}
+ \ln\left(\rho \right) \Biggl[\frac{191 }{27}  +\frac{8 }{3}  \Li_{2}(x)  \Biggr] + \frac{40}{9}\Li_{2}(x)     + \frac{1165}{81}
\Biggr\}
-\ln\left(\rho\right) \Biggl[
\nn\\
& & \hspace{10mm}
+ \frac{32 x^4-46 x^3+33 x^2+8 x-4}{6 x^2} \zeta(2)
-\frac{\left(x^2-x+1\right)
\left(4 x^2-7 x+4\right)}{6 x^2} \ln(1 - x)^2
\nn\\
& & \hspace{10mm}
-\frac{20 x^4 \! - \! 31 x^3 \! + \! 60 x^2 \! - \! 31 x \! + \! 20}{18 x^2}\ln(1 - x)
+ \frac{20 x^4 \! - \! 67 x^3 \! + \! 141 x^2 \! - \! 112 x \! + \! 74}{18 x^2} \ln(x)  \nn\\
& & \hspace{10mm}
+ \frac{8 x^4-x^3-15 x^2+17 x-4}{12 x^2} \ln(x)^2
- \frac{(2 x-1) \left(4 x^3-3 x^2+4\right)}{6 x^2} \ln(x) \ln(1 - x)
\Biggr]  \nn\\
& & \hspace{10mm}
+ \frac{(2 x-1) \left(x^2-x+1\right)}{3 x} \zeta(3)
- \frac{(x-1)^2 \left(x^2-x+1\right)}{9 x^2}\ln^3(1 - x)  \nn \\
& & \hspace{10mm}
-  \frac{196 x^4 \! - \! 311 x^3 \! + \! 258 x^2 \! + \! 13 x \! - \! 38}{18 x^2} \zeta(2)
- \frac{2 \left(2 x^4 \! -\! 9 x^3\! +\! 16 x^2\! -\! 11 x\! +\!
4\right)}{3 x^2} \ln(1 \! - \! x) \zeta(2)  \nn\\
& & \hspace{10mm}
+ \frac{12 x^4 \! - \! 20 x^3 \! - \! x^2 \! + \! 24 x \! - \! 4}{6 x^2}
\ln(x) \zeta(2)
+ \frac{2 (1-x^2) \left(x^2-x+1\right)}{3 x^2} \ln(1 - x)
\Li_{2}(x)  \nn\\
& & \hspace{10mm}
+ \frac{7 \left(16 x^4 \! - \! 23 x^3 \! + \! 48 x^2 \! - \! 23 x \! + \! 16\right)}{54 x^2}
\ln(1  \! -  \! x)   \!
+  \! \frac{20 x^4 \! - \! 58 x^3 \! + \! 81 x^2 \! - \! 58 x \! + \! 20}{18 x^2}
\ln^2(1  \! -  \! x)  \nn\\
& & \hspace{10mm}
- \frac{4 x^3 \! - \! 6 x^2 \! + \! 7 x \! - \! 4}{12 x}\ln(x) \ln^2(1  \! -  \! x)
+ \frac{40 x^4 \! - \! 50 x^3 \! + \! 9 x^2 \! + \! 37 x \! - \! 20}{18 x^2}\ln(x) \ln(1  \! -  \! x)   \nn\\
& & \hspace{10mm}
- \frac{x^4-3 x^3+4 x^2-x+1}{3 x^2}
\ln^2(x) \ln(1 - x)
 +\frac{4 x^4-2 x^3-22 x^2+31 x-4}{36 x^2} \ln^3(x)    \nn\\
& & \hspace{10mm}
- \frac{20 x^4+8 x^3-84 x^2+92 x-55}{18 x^2}\! \ln^2(x)
 -\frac{\left(x^2-x+1\right) \!\left(2 x^2-7 x+12\right)}{3 x^2}\!
 \ln(x) \Li_{2}(x)\nn\\
& & \hspace{10mm}
- \frac{112 x^4-449 x^3+1011 x^2-836 x+562}{54 x^2} \ln(x)   \nn\\
& & \hspace{10mm}
+ \frac{2 (1-x^2) \left(x^2-x+1\right)}{3 x^2}\Li_{3}(1 - x)
+\frac{\left(x^2-x+1\right) \left(2 x^2-3 x+4\right)}{3 x^2} \Li_{3}(x) \nn\\
& & \hspace{10mm} -\left(Q_f^2-1\right)\frac{(1-x+x^2)}{x^2}\Biggl[
(1\!-\!x\!+\!x^2)\left(\frac{5}{12} \!-\!2 \zeta(3) \!+\! \frac{1}{2}\ln\left(\rho\right) \right) - \frac{2-x}{4}
\ln\left(x\right)\Biggr] 
\, , \label{f0}
\end{eqnarray}
where $Q_f$ is the charge of the heavy fermion in units of the electron
charge. Eq.~(\ref{f0}) is in agreement with the result of 
Refs.~\cite{BecMel,Act}. The next-to-leading term is new and reads
\begin{eqnarray}
& & \hspace{-5mm} f_1(\rho,x) = \frac{2 (x-1) \left(x^2-x+1\right)
\left(2 x^2+x+2\right)}{x^3} \Biggl[
\ln^2\left(\rho\right) +4 \Li_{2}(x)  + 12 \Biggr] \nn \\
& &  \hspace{8mm} + \ln\left(\rho\right) \Biggl\{
  \frac{(x-1)}{x^3} \left(2 x^4 \! - \! 5 x^3 \! + \! 5 x^2 \! - \! 5 x \! + \! 2\right)
- \frac{2 \left(x^2 \! - \! x \! + \! 1\right)}{ x^3} \Bigl[
 \left( 2 x^3 \! -  \! x^2 \! +  \! 2 x \! - \! 4 \right) \ln(x)  \nn \\
& &  \hspace{8mm} - \left( x-1\right) \left(2 x^2+ x+2\right) \ln(1
- x) \Bigr] \Biggr\}
- \frac{40 x^5-54 x^4+50 x^3-17 x^2-12 x+8}{x^3}\zeta(2)   \nn \\
& &  \hspace{8mm} + \frac{(x \! - \! 1)}{2 x^3} \Bigl[ 2 \bigl(12
x^4 \! - \! 5 x^3 \! + \! 13 x^2 \! - \! 5 x \! + \! 12 \bigr) +
(8x^4 \! - \! 6x^3 \! + \! 9x^2 \! - \! 6x \!
+ \! 8) \ln (1  \! -  \! x) \Bigr]\ln (1  \! -  \! x) \nn \\
& &  \hspace{8mm} - \biggl[ \frac{12 x^5 \! - \! 21 x^4 \! + \! 26
x^3 \! - \! 26 x^2 \! + \! 21 x \! - \! 14}{x^3} - \frac{8x^5 \! -
\! 10x^4 \! + \! 8x^3 \! - \! x^2 \! - \! 6x \!
+ \! 4}{x^3} \ln(1  \! -  \! x)  \nn \\
& &  \hspace{8mm} + \frac{8x^5-7x^4-x^3+12x^2-15x+8}{2 x^3} \ln(x)
\biggr] \ln(x) + \frac{3\left(Q_f^2-1\right)}{x^3} \Bigl[ (2 \!-\! 3x\! +\! 4x^2\! -\! 4x^3 
\nn \\
& & \hspace{8mm} 
+ \!3x^4 \!-\!
2x^5) \ln\left(\rho \right) - \left(2 - 3 x + 3 x^2-
x^3\right) \ln\left( x\right)
\Bigr] \, . \label{f1}
\end{eqnarray}
The expansion in the large-mass limit  takes the form
\begin{equation}
f(\rho,x)=\sum_{n=0}^\infty\rho^{-n}\bar{f}_n(\rho,x)\,,
\label{lmexp}
\end{equation}
where the leading $n=0$ term vanishes because of the renormalization
condition and $\bar{f}_n(\rho,x)$ depend on $\rho$ only logarithmically.
For the next-to-leading term we obtain
\begin{eqnarray}
\bar{f}_1(\rho,x)&=&
\frac{955 x^3-3926 x^2+3926 x-955}{1350 x}
- \frac{12 x^3-19 x^2+14 x-6}{10 x}\zeta(2)
\nn \\
& &    +     \frac{3 x^3+x^2-x-3}{30 x}\ln{(1 - x)}
 +    \frac{2 x^3-5 x^2+5 x-2}{20 x}   \ln^2 {(1 - x)}
\nn \\
& &
+  \frac{5 x^3-22 x^2+22 x-5}{30 x}  \ln\left(\rho \right)
-  \frac{20 x^3-78 x^2+93 x-58}{90 x}  \ln{x}
\nn \\
& &    + \frac{12 x^3-19 x^2+14 x-6}{30 x}  \ln{x}\ln{(1 - x)}
-  \frac{1}{60} \left(6 x^2-x-4\right) \ln^2 {x}
\nn \\
& &   +   \frac{4 \left(x^3-2 x^2+2 x-1\right)}{5 x}  \mbox{Li}_2(x) 
-\left(Q_f^2-1\right) \frac{41(x^3 -2 x^2 + 2 x -1)}{54
x } \,.
\label{f1bar}
\end{eqnarray}


\end{document}